\documentclass[12pt]{article}
\usepackage{amsmath}
\usepackage{amssymb}
\usepackage{amsfonts}
\newcommand{\be}{\begin{equation}}
\newcommand{\ba}{\begin{eqnarray}}
\newcommand{\ee}{\end{equation}}
\newcommand{\ea}{\end{eqnarray}}
\newcommand{\cosech} { {\rm cosech}}
\DeclareMathOperator{\sech}{sech}

\newcommand{\ket}[1]{\ensuremath{\left|#1\right\rangle}}

\begin{document}

\title{Parametric symmetries in exactly solvable real and $PT$ symmetric complex potentials}
\author{Rajesh Kumar Yadav$^{a}$\footnote{e-mail address: rajeshastrophysics@gmail.com (R.K.Y)}, Avinash Khare$^{b}$\footnote {e-mail address: khare@physics.unipune.ac.in (A.K)}, Bijan Bagchi$^{c}$\footnote{e-mail: bbagchi123@gmail.com (B.B)}, Nisha Kumari$^{d}$\footnote{e-mail address: nishaism0086@gmail.com (N.K)} and \\ Bhabani Prasad Mandal$^{d}$\footnote{e-mail address: bhabani.mandal@gmail.com (B.P.M).}}
 \maketitle
 {$~^a$ Department of Physics, S. P. College, S K. M. University, Dumka-814101, India.\\
 $~^b$ Department of Physics, Savitribai Phule Pune University, Pune-411007, India.\\
 $~^c$ Department of Applied Mathematics, University of Calcutta, Kolkata-700 009, India.\\
$~^d$ Department of Physics, Banaras Hindu University, Varanasi-221005, India. 
}

\begin{abstract}

In this paper, we discuss the parametric symmetries in different exactly solvable systems characterized by real or
complex $PT$ symmetric potentials. We focus our attention on the conventional potentials
such as the generalized P\"oschl Teller (GPT), Scarf-I and $PT$ symmetric
Scarf-II which are invariant under certain parametric transformations. The resulting set of potentials are 
shown to yield a completely different behavior of the bound state solutions. Further the supersymmetric (SUSY)
partner potentials acquire different forms under such parametric transformations leading to  new sets of 
exactly solvable real and $PT$ symmetric complex potentials. These potentials are also observed to be 
 shape invariant (SI) in nature. We subsequently take up a study of the newly discovered rationally extended SI Potentials,
corresponding to the above mentioned conventional potentials, whose bound  state solutions are
associated with the exceptional orthogonal polynomials (EOPs). We discuss the transformations of the corresponding Casimir operator 
employing the properties of the $so(2,1)$ algebra.
\end{abstract}

\section{Introduction}

The exactly solvable (ES) models play an important role in our understanding of the bound state problems in quantum mechanics.  However, in the literature, ES systems are hard come by as is evidenced by the appearence of only a handful of potentials that yield to an exact treatment \cite{fluge,cks}. With the advent of supersymmetric quantum mechanics (SUSYQM), following a remarkable paper by Witten  in $1981$ \cite{wit}, it was realized that SUSY offers a clue to the general nature of solvability which basically amounts to a process of factorizing the Schr\"odinger Hamiltonian. In effect this means that a nonlinear differential equation for the superpotential that belongs to a Riccati's class needs to be solved  and that it is only for a limited choices of the superpotentials that such a criterion can be fulfilled. From the superpotential it is always possible to work out a pair of isospectral partner Hamiltonians satisfying the condition of shape invariance (SI) \cite{gen}. In the unbroken case of SUSY to which we shall restrict ourselves here, the ground state is nondegenerate but otherwise both the Hamiltonians have an identical column of energies except with the ground state belonging to only one of the components but not to both.

The list of exactly solvable systems was further expanded in the context of polynomial Heisenberg algebras \cite{fh, cfnn} which offer additional degeneracies of the energy levels
and new  families exceptional orthogonal polynomials (EOPs) (also known as $X_m$ Laguerre and $X_m$ Jacobi
polynomials) \cite{dnr1,dnr2,xm2,xm1, qn, mq} that start with polynomials of degree one or higher but could be reduced to tractable forms of differential equations which give solvable forms of the spectra \cite {que,bqr,os,dim,hj,pdm,fplank, nfold1,nfold2,qscat1, qscat2, qscat3, tdse1, tdse2, tdse3,tdse4}. These include three new classes of exactly  solvable and $SI$ potentials \cite{que,bqr,os,dim,gtextd} which could be identified as the rational extensions of the radial oscillator, generalized  P\"oschl Teller (GPT) and Scarf I potentials.

A parallel development concerning complex extensions of quantum mechanics, a subclass of which is controlled by an underlying combined parity (P) and time reversal (T) symmetry  that yields a large family
of exactly solvable potentials, have also attracted much attention over the past one and half decades due to the realization of
fully consistent
quantum theories within such a framework \cite {bender, ben, mos}. In the present work one of our aim is to discuss the parametric symmetries for some of
the exactly solvable Hermitian as well as
PT symmetric complex potentials and their rational extensions counterparts. For concreteness we would focus on the conventional GPT, Scarf-I and $PT$ symmetric
Scarf-II potentials.
We observe that while under certain  parametric transformations these  potentials remain invariant, the  set of bound states change
 to a different one. In fact, within $SUSYQM$, the corresponding superpotentials for these potentials change under such parametric transformation leading to new partner potentials which satisfy  usual $SI$ property. Thus the parametric symmetry in these potentials is responsible for a
previously unnoticed set of bound state solutions as well as generating new ES potentials. We further apply the same technique to rationally extended GPT, Scarf-I and PT symmetric Scarf-II potentials and
obtain the new bound state solutions associated with these potentials. In all three cases it turns out possible for the new solutions to be written in
 terms of EOPs. This result is of interest.
We further  observe that the
scattering amplitudes corresponding to the two sets of bound states for GPT and complex $PT$ symmetric Scarf II potentials
remain invariant under these parametric transformations. We also determine new ES potentials which are isospectral to the conventional
potentials for all three cases.

A point to note here: An algebraic technique based upon the use of $so(2,1)$ algebra for the Schr\"odinger equation to construct exactly solvable potentials in quantum mechanics was pioneered long time ago by Alhassid et al \cite{potalg1,potalg2,potalg3, potalg4, potalg5} along with some other groups \cite{potalg6, potalg7,cvs} in a series of papers which were later extended in \cite{eq} by considering more general possibilities for the generators. Subsequently complex extensions of these works were carried out in \cite{sl2c, pt1,pt2} and interesting results were derived such as the existence of two series of energy levels stemming from two noncommuting classes of complex Lie algebras. In particular for the $PT$-symmetric complex Scarf-II potential it was observed that it supports two series of real eigenvalues with physically acceptable wavefunctions that is related to the invariance of the potential under exchange of its coupling parameters.

These group theoretic techniques were recently extended to the case of rationally extended potentials by extending the generators
of the associated $so(2,1)$ group through the introduction of a new operator $U$ to express the Hamiltonian in terms of Casimir of the group. Hence the bound
states corresponding to the rationally extended potentials are obtained in
terms of EOPs in an elegant fashion \cite{gtextd}. In the following we will discuss how the generators get
modified under this parametric symmetry of the potentials .

The plan of the present paper is as follows:
In section $2$, we discuss the parametric symmetries in conventional GPT, Scarf-I and
$PT$ symmetric complex Scarf-II potentials and obtained
their new solutions with new superpotentials. Corresponding to these  conventional potentials
new rationally extended potentials whose solutions are in terms of EOPs are also obtain in this section.
In section $3$, we discuss briefly the $so(2,1)$ algebra corresponding to these extended potentials and obtain
the modified generators of rationally extended GPT potential. The algebra corresponding to the rationally extended
$PT$ symmetric Scarf II  (i.e. $sl(2,\mathbb{C})$ algebra) and  the rationally extended Scarf I (i.e $iso(2,1)$ algebra)
potentials are discussed briefly. The corresponding modified generators are also obtained in this section.
Finally we summarize the results obtained in Section $4$.

\section{New bound state solutions of  exactly solvable conventional potentials and their rationally extended counterparts}

In this section, we focus on few exactly solvable shape invariant potentials which are invariant
under certain parametric transformations. The curious thing is that the forms of their supersymmtric partner potentials change under such transformations. This allows us to obtain previously unnoticed sets of bound state solutions of the conventional potentials. We have based our investigations on three different types of potentials, namely the  generalized
P\"oschl Teller (GPT) potential, trigonometric Scarf (Scarf I) and $PT$ symmetric complex Scarf II potentials. In what to follow the new potentials determined by us will be used as a springboard to undertake calculations for the more complicated rational extensions of such potentials and go for a potential algebra treatment to tie them up in a single framework of $so(2,1)$ complex algebra. The rationally extended SI potentials and their solutions in terms of EOPs corresponding to the conventional
GPT, Scarf I and $PT$ symmetric complex Scarf II potentials are already obtained in Refs. \cite{que,bqr}. We shall briefly review certain aspects of the existing solutions and then proceed to obtain new sets of rationally extended real and $PT$ symmetric complex SI potentials
corresponding to the above three new conventional potentials by using  parametric transformation which leaves the usual rationally extended potentials invariant.

\subsection{GPT potential}
The conventional GPT potential defined on the half-line $0 <x < \infty$ is given by
\be\label{gpt1}
V^{(A,B)}_{1,GPT}(x)=(B^2+A(A+1))\cosech^2 x-B(2A+1)\cosech x \coth x.
\ee
The accompanying bound state energy eigenvalues and the eigenfunctions are \cite{cks}
\be\label{gptee1}
E_n=-(A-n)^2;\qquad n=0,1,2.......n_{\mbox{max}}<A
\ee
and
\be\label{gptwf1}
\psi^{(A,B)}_n(x)=N_n (\cosh x-1)^{\frac{B-A}{2}} (\cosh x+1)^{-\frac{B+A}{2}} P^{(\alpha,\beta)}_n(\cosh x), \quad B>A+1>1,
\ee
where $\alpha=B-A-\frac{1}{2}$, $\beta=-B-A-\frac{1}{2}$ and $N_n$ is the normalization constant.

Corresponding to (\ref{gpt1}) the superpotential $W(x)$ is known to be
\be
W(x)=A\coth x-B\cosech x.
\ee
As a result the partner potential to the GPT potential can be written as
\ba
V^{(A,B)}_{2,GPT}(x)&=&W^2(x)+\frac{dW(x)}{dx}\nonumber\\
&=& (B^2+A(A-1))\cosech^2 x-B(2A-1)\cosech x \coth x.
\ea
The potential $V^{(A,B)}_{2,GPT}(x)$ can be recognized to be  shape invariant \cite{cks} under a simple translation  $A\rightarrow A-1$.

The scattering amplitude corresponding to (\ref{gpt1}) has the form \cite{ks}
\be\label{scatt_gpt}
 S_{l=0}(k)= 2^{-4ik}\frac{\Gamma(2ik)\Gamma(-A-ik)\Gamma(B+\frac{1}{2}-ik) }
{\Gamma(-2ik)\Gamma(-A+ik) \Gamma(B+\frac{1}{2}+ik)}.
\ee
where $k$ is the wave number. The poles of the gamma functions give the bound state energy spectrum (\ref{gptee1}).

We next exploit an interesting property of $V^{(A,B)}_{1,GPT}(x)$ that its form is unaffected under the joint replacements of $B \longleftrightarrow (A+\frac{1}{2})$ in addition to being symmetric corresponding to simultaneous transformations $B\longrightarrow -B$ and  $(A+\frac{1}{2}) \longrightarrow -( A+\frac{1}{2})$. From the consideration of the correct asymptotic behavior of the bound state wave functions we can restrict, without loss of generality, to $B > 0$ , $A > -\frac{1}{2}$. If we apply the former transformations of parameters we notice that the energy eigenvalues and the eigenfunctions of (\ref{gpt1}) acquire completely different  forms being given by
\be\label{gptee2}
E_n=-(B-n-\frac{1}{2})^2;\qquad n=0,1,2.......n_{\mbox{max}}<B-\frac{1}{2}
\ee
and
\be\label{gptwf2}
\psi^{(B\leftrightarrow A+\frac{1}{2})}_n(x)=N^{(B\leftrightarrow A+1/2)}_n (\cosh x-1)^{\frac{A-B+1}{2}} (\cosh x+1)^{-\frac{B+A}{2}} P^{(\alpha,\beta)}_n(\cosh x),   
\ee
with $A>-\frac{1}{2}, B>0$ and the parameters $\alpha=A-B+\frac{1}{2}$, $\beta=-A-B-\frac{1}{2}$. The transposition $B\longleftrightarrow  (A+\frac{1}{2})$ also leads to a different but a perfectly acceptable superpotential to (\ref{gpt1}) namely
\be\label{spgpt1}
W(x)=(B-\frac{1}{2})\coth x-(A+\frac{1}{2})\cosech x.
\ee
It induces a partner potential that has the form
\be
V^{(A,B)}_{2,GPT}(x)=((B-1)^2+A(A+1))\cosech^2 x-(B-1)(2A+1)\cosech x \coth x.
\ee
One can notice that (\ref{gpt1}) is also shape invariant under the translation $B\rightarrow B-1$.

The scattering amplitudes for the $s$-wave ($l=0$) to include the new state of bound states (\ref{gptee2}) is obtained from (\ref{scatt_gpt}) by making the replacements $B \longleftrightarrow (A+\frac{1}{2})$:
\be\label{smatrix}
 S^{(B\leftrightarrow A+\frac{1}{2})}_{l=0}(k)= 2^{-4ik}\frac{\Gamma(2ik)\Gamma(-B+\frac{1}{2}-ik)\Gamma(A-ik+1) }
{\Gamma(-2ik)\Gamma(-B+\frac{1}{2}+ik) \Gamma(A+ik+1)}.
\ee
Thus we see that the $GPT$ potential has another scattering matrix\footnote{Unlike the above two parametric transformations
 transformations i.e $B\longleftrightarrow (A+\frac{1}{2})$ and $B\longrightarrow -B, (A+\frac{1}{2})\longrightarrow -(A+\frac{1}{2})$,
 third parametric transformation $B\longleftrightarrow -(A+\frac{1}{2})$ is also possible under which the S-matrix (\ref{scatt_gpt}) remains invariant.} because of its invariance under the interplay of its coupling parameters. The poles of S-matrix (\ref{smatrix}) giving correct bound states (\ref{gptee2}). Here
we notice that thus so long as $B-A-\frac{1}{2}$ is not an integer, GPT has two sets of nodeless states, two states with one nodes etc. The true
ground state of the system will depend on which of $A$ and $B-\frac{1}{2}$ is less.

\subsubsection{Rationally extended GPT potential}

The rationally extended SI GPT potential \cite{bqr} which is isospectral to the conventional GPT potential
(\ref{gpt1}), defined for the parameter $B>A+1>1$ is given by
\ba\label{gptex}
V^{(A,B)}_{1,extd}(x)&=&(B^2+A(A+1))\cosech^2 x-B(2A+1) \cosech x \coth x \nonumber\\
&+& \frac{2(2A+1)}{2B\cosh x-2A-1}-\frac{2(4B^2-(2A+1)^2)}{(2B\cosh x-2A-1)^2}.
\ea
 The wave functions of this extended
potential in terms of $X_1$ exceptional Jacobi polynomials, $\hat{P}^{(\alpha,\beta)}_{n}(\cosh x)$ is given by
\be
\psi^{(A,B)}_{n,extd} (x)=N_{n,extd}\times \frac{(\cosh x-1)^{\frac{(B-A)}{2}}(\cosh x+1)^{-\frac{(B+A)}{2}}}{[2B\cosh x-2A-1]}\hat{P}^{(\alpha,\beta)}_{n+1}(\cosh x),
\ee
where the parameters $\alpha$ and $\beta$ are same as defined in Eq. (\ref{gptwf1}) and $N_{n,extd}$ is the normalization constant.
The superpotential corresponding to this potential is given by
\ba
W(x)&=&A\coth x-B\cosech x+2B\sinh x\nonumber\\
&\times &\bigg(\frac{1}{(2B\cosh x-2A-1)}-\frac{1}{(2B\cosh x-2A+1)}\bigg).
\ea
 On changing the parameters $B\longleftrightarrow (A+\frac{1}{2})$, the above extended potential (\ref{gptex}) becomes
 \ba\label{gptex1}
V^{(A,B)}_{1,extd}(x)&=&(B^2+A(A+1))\cosech^2 x-B(2A+1) \cosech x \coth x \nonumber\\
&+& \frac{4B}{2(A+\frac{1}{2})\cosh x-2B}-\frac{8((A+\frac{1}{2})^2-B^2)}{(2(A+\frac{1}{2})\cosh x-2B)^2},
\ea
and hence is not invariant unlike the conventional GPT potential (\ref{gpt1}). This new rationally extended GPT potential
is isospectral to the new conventional GPT potential whose bound state spectrums are given in (\ref{gptee2}).
 The wave functions of this new extended
potential in terms of $X_1$ exceptional Jacobi polynomials, $\hat{P}^{(\alpha,\beta)}_{n}(\cosh x)$ is given by
\be
\psi^{(B\leftrightarrow A+\frac{1}{2})}_{n,extd} (x)=N^{B\leftrightarrow A+\frac{1}{2}}_{n,extd}\times \frac{(\cosh x-1)^{\frac{A-B+1}{2}}(\cosh x+1)^{-\frac{B+A}{2}}}{[2(A+\frac{1}{2})\cosh x-2B]}\hat{P}^{(\alpha,\beta)}_{n+1}(\cosh x).
\ee
The superpotential is given by
\ba
W(x)&=&(B-\frac{1}{2})\coth x-(A+\frac{1}{2})\cosech x+2(A+\frac{1}{2})\sinh x\nonumber\\
&\times &\bigg(\frac{1}{(2(A+\frac{1}{2})\cosh x-2B)}-\frac{1}{(2(A+\frac{1}{2})\cosh x-2B+2)}\bigg).
\ea
Using this superpotential, we get the partner potential
\ba
V^{(A,B)}_{2,extd}(x)&=&((B-1)^2+A(A+1))\cosech^2 x-(B-1)(2A+1)\cosech x \coth x\nonumber\\
&+&\frac{4(B-1)}{[2(A+\frac{1}{2})\cosh x-2B+2]}-\frac{8((A+\frac{1}{2})^2-(B-1)^2)}{[2(A+\frac{1}{2})\cosh x-2B+2]^2}.
\ea
It can be easily shown that this new rationally extended GPT potential is also shape invariant under the translation
of parameter $B\rightarrow B-1$.

 The above extended SI GPT potential (\ref{gptex1}) isospectral to (\ref{gptee2}) is easily generalized to the $X_m$ case given by
\ba\label{gptpot3}
V^{(A,B)}_{m,extd}(x)&=&(B^2+A(A+1))  \cosech^2 \ x-B(2A+1)  \cosech \ x  \coth x\nonumber\\
&+&2m(2A-m+2)-(2A-m+2)[(2B-2(A+1)\cosh x)]\nonumber\\
&\times & \frac{P_{m-1}^{(-\alpha,\beta)}(\cosh x)}{P_{m}^{(-\alpha-1,\beta-1)}(\cosh x)}+\frac{(2A-m+2)^2\sinh^2 x}{2}\nonumber\\
&\times &\bigg(\frac{P_{m-1}^{(-\alpha,\beta)}(\cosh x)}{P_{m}^{(-\alpha-1,\beta-1)}(\cosh x)}\bigg)^2 ; \quad 0\leq x\leq \infty.\nonumber\\
\ea
The corresponding wavefunctions in terms of $X_m$ Jacobi polynomials $(\hat{P}_{n+m}^{(\alpha,\beta )}(\cosh x))$ are given by
\be\label{wfgpt3}
\psi^{(B\leftrightarrow A+\frac{1}{2})}_{n,m}(x) = N^{(B\leftrightarrow A+\frac{1}{2})}_{n,m,extd}\times \frac{(\cosh x-1)^{(\frac{A-B+1}{2})}(\cosh x+1)^{-(\frac{B+A}{2})}}{P_{m}^{(-B-A-\frac{1}{2},-B-A-\frac{3}{2})}(\cosh x)}\hat{P}_{n+m}^{(\alpha,\beta)}(\cosh x). 
\ee
 The scattering amplitudes corresponding to the above potentials (\ref{gptpot3}) are obtained by taking the asymptotic behaviors of 
the associated $X_m$ Jacobi polynomials given by
 \ba\label{sctm}
S^m_{l=0} &=& S^{(B\leftrightarrow A+\frac{1}{2})}_{l=0}(k)\left[\frac{\{ (A+\frac{1}{2})^2-(ik-\frac{1}{2})^2 \} + (A-ik+1)(1-m)}{\{ (A+\frac{1}{2})^2-(ik+\frac{1}{2})^2\} +(A+ik+1)(1-m)}\right]. 
\ea
For $m=1$, the scattering amplitude corresponds to the $X_1$ case and in the limit $m=0$ it reduces to $S^{(B\leftrightarrow A+\frac{1}{2})}_{l=0}(k)$ 
given in Eq. (\ref{smatrix}).
The above scattering amplitudes (\ref{sctm}) can be also obtained simply by replacing $B\longleftrightarrow A+\frac{1}{2}$ in the scattering amplitudes 
obtained in Ref. \cite{qscat2}.

\subsection{Trigonometric Scarf or Scarf I potential}

We now consider the second example namely, the conventional Scarf I potential which reads
in the standard form
\be\label{scarf1}
V_{1,Scarf}^{(A,B)}(x)=(B^2+A(A-1))\sec^2 x-B(2A-1)\sec x \tan x; \qquad -\frac{\pi}{2}<x<\frac{\pi}{2}.
\ee
The energy eigenvalues and eigenfunctions are \cite{cks}
\be
E_n=(A+n)^2;\qquad n=0,1,2....
\ee
and
\be
\psi^{(A,B)}_n(x)=N_n (1-\sin x)^{\frac{A-B}{2}}(1+\sin x)^{\frac{A+B}{2}}P^{(\alpha, \beta)}_n(\sin x), \quad 0<B<A-1,
\ee
where  $\alpha=A-B-\frac{1}{2}$, $\beta=A+B-\frac{1}{2}$ and $N_n$ is the normalization constant.
The superpotential corresponding to this potential is already known and given by
\be
W(x)=A\tan x-B\sec x,
\ee
yielding the partner potential
\be\label{scarfpartner}
V^{(A,B)}_{2,Scarf}(x) = (B^2+A(A+1))\sec^2 x-B(2A+1)\sec x \tan x.
\ee
This potential is shape invariant under the translation of parameter $A\rightarrow A+1$.

On transforming the parameters $B\longleftrightarrow (A-\frac{1}{2})$, the given Scarf I potential (\ref{scarf1}) remains invariant,
but the energy eigenvalues, eigenfunctions and the superpotential have completely different forms and are given by
\be\label{scarfee1}
E_n=(B+n+\frac{1}{2})^2; \qquad n=0,1,2.....
\ee
\be\label{scarfwf1}
\psi^{(B\leftrightarrow A-\frac{1}{2})}_n(x)=N^{B\leftrightarrow A-\frac{1}{2}}_n (1-\sin x)^{\frac{B-A+1}{2}}(1+\sin x)^{\frac{A+B}{2}}P^{(\alpha, \beta)}_n(\sin x); \quad B>A-1>0,
\ee
and
\be
W(x)=(B+\frac{1}{2})\tan x-(A-\frac{1}{2})\sec x,
\ee
with a new set of parameters $\alpha=B-A+\frac{1}{2}$ and $\beta=A+B-\frac{1}{2}$.\\
The partner potential corresponding to this new system is given by
\be
V^{(A,B)}_{2,Scarf}(x)=((B+1)^2+A(A-1))\sec^2 x-(B+1)(2A-1)\sec x \tan x.
\ee
Thus, we observe that on changing the parameters, the invariance potential $V_1(x)$ under $B\leftrightarrow (A-\frac{1}{2})$ allows, as in the previous case of the $GPT$ potential,
a new defining superpotential which provides a different  partner potential to the Scarf-I potential than the one considered in (\ref{scarfpartner}).  The latter
is also SI under the translation of a different  parameter $B\longrightarrow B+1$. Here we also notice that the parametric transformation $B\leftrightarrow (A-\frac{1}{2})$ generates two sets of bound states for the Scarf I potentials. In other words we have two sets of nodeless states, two states with one
node etc, but true ground state of the system will depends on which of $A$ and $B+\frac{1}{2}$ is less.

\subsubsection {Rationally extended Scarf I potential}

The rationally extended Scarf I potential \cite{que} isospectral to the conventional one (\ref{scarf1}) is given by
\ba\label{extsc_con_pot}
V^{(A,B)}_{1, extd}(x)&=&(B^2+A(A-1))\sec^2 x-B(2A-1)\sec x \tan x\nonumber\\
&+& \frac{2(2A-1)}{(2A-1-2B\sin x)}-\frac{2[(2A-1)^2-B^2]}{(2A-1-2B\sin x)^2},\quad 0<B<A-1.
\ea
The wavefunctions in terms of exceptional $X_1$ Jacobi polynomials are 
\be\label{exsc_con_wf}
\psi^{(A,B)}_n(x)=N_{n,extd}\times\frac{(1-\sin x)^{\frac{A-B}{2}}(1+\sin x)^{\frac{A+B}{2}}}{(2A-1-2B\sin x)}\hat{P}^{(\alpha,\beta)}_{n+1}(\sin x).
\ee
The generalization to the $X_m$ case is well known and given in detail in Ref. \cite{os}.

In the case of conventional Scarf I potential (\ref{scarf1}), we have shown that on changing the parameters $(B\leftrightarrow A-\frac{1}{2})$, the
potential is remain invariant but the bound states solutions are different which are
given in Eqs. (\ref{scarfee1}) and (\ref{scarfwf1}). Under this parametric transformation the rationally extended potential (\ref{extsc_con_pot})
 is not invariant and we get a new rationally extended Scarf I potential isospectral to the conventional Scarf I potential (\ref{scarf1}) 
with the energy eigenvalues (\ref{scarfee1}). This new set of rationally extended Scarf I potential is given by
\ba\label{nscarf}
V^{(A,B)}_{1, extd}(x)&=&(B^2+A(A-1))\sec^2 x-B(2A-1)\sec x \tan x\nonumber\\
&+& \frac{4B}{(2B-2(A-\frac{1}{2})\sin x)}-\frac{8(B^2-(A-\frac{1}{2})^2)}{(2B-2(A-\frac{1}{2})\sin x)^2}; 
\ea
with $B>A-1>0$. The wavefunctions of this potential in terms of $X_1$ exceptional Jacobi orthogonal polynomials, $\hat{P}^{(\alpha,\beta)}_n(\sin x)$
become
\be
\psi^{(B\leftrightarrow A-\frac{1}{2})}_n(x)=N^{B\leftrightarrow A-\frac{1}{2}}_{n,ext}\times\frac{(1-\sin x)^{\frac{B-A+1}{2}}(1+\sin x)^{\frac{A+B}{2}}}{2B-2(A-\frac{1}{2})\sin x}\hat{P}^{(\alpha,\beta)}_{n+1}(\sin x).
\ee
The superpotential for this new rationally extended Scarf I potential is
\ba
W(x)&=&(B+\frac{1}{2})\tan x-(A-\frac{1}{2})\sec x+2(A-\frac{1}{2})\cos x\nonumber\\
&\times &\bigg(\frac{1}{2B+2-(2A-1)\sin x}-\frac{1}{2B-(2A-1)\sin x}\bigg).
\ea
Thus the partner potential can easily be obtained and is given by
\ba
V^{(A,B)}_{2, extd}(x)&=&((B+1)^2+A(A-1))\sec^2 x-(B+1)(2A-1)\sec x \tan x\nonumber\\
&+& \frac{4(B+1)}{(2(B+1)-2(A-\frac{1}{2})\sin x)}-\frac{8((B+1)^2-(A-\frac{1}{2})^2)}{(2(B+1)-2(A-\frac{1}{2})\sin x)^2}.
\ea
This new rationally extended potential is also translationally SI under the translation of parameter $B\longrightarrow B+1$.\\
The potentials corresponding to the $X_m$ case and their wavefunctions are given by 
\ba\label{nscarfxm}
V^{(A,B)}_{m,extd}(x)&=&\frac{(2\alpha^2+2\beta^2)}{4}\sec^2 x-\frac{(\beta^2-\alpha^2)}{2}\sec x \tan x-2m(\alpha-\beta-m+1)\nonumber\\
&-&(\alpha -\beta -m+1)(\alpha +\beta +(\alpha -\beta +1)\sin x)\frac{P^{(-\alpha,\beta )}_{m-1}(\sin x)}{P^{(-\alpha-1,\beta-1 )}_{m}(\sin x)}\nonumber\\
&+&\frac{(\alpha -\beta -m+1)^2\cos^2 x}{2}\bigg(\frac{P^{(-\alpha,\beta )}_{m-1}(\sin x)}{P^{(-\alpha-1,\beta-1 )}_{m}(\sin x)}\bigg)^2, \quad -\frac{\pi}{2}<x<\frac{\pi}{2}\nonumber\\
\ea
and
\be\label{scm3}
\psi^{(A,B)}_{n,m}(x) = N_{n,m,extd}\times \frac{(1-\sin x)^{\frac{1}{2}(\alpha +\frac{1}{2})}(1+\sin x)^{\frac{1}{2}(\beta +\frac{1}{2})}}{P^{(-\alpha-1, \beta-1 )}_m(\sin x)} \hat{P}^{(\alpha,\beta)}_{n+m}(\sin x) 
\ee
respectively. The above potentials are isospectral to their conventional one (i.e the energy eigenvalues are same as given in (\ref{scarfee1}))
and are also SI under the translation of parameter $B\longrightarrow B+1$.

\subsection{$PT$ symmetric complex Scarf-II potential}

As a third example, we consider the well known complex and $PT$ symmetric Scarf II potential  \cite{sl2c, pt1, pt2, ap, ply, za,glpt}. In the usual notations the $PT$ symmetric
Scarf II potential defined on the full line $-\infty<x<\infty$ is given by
\be\label{scarfii_pt}
V^{(A,B)}_{1,Scarf II}(x)=-(B^2+A(A+1))\sech^2 x+iB(2A+1)\sech x\tanh x;\qquad A>B-\frac{1}{2} >0.
\ee
With the above definition of $V^{(A,B)}_{1,Scarf II}(x)$, the energy eigenvalues are real \cite{pt1} namely,
\be\label{scarfptee1}
E_n=-(A-n)^2;\qquad n=0,1,2....n_{\mbox{max}}<A,
\ee
and the accompanying eigenfunctions are given by
\be
\psi^{(A,B)}_n(x)=N_n (\sech x)^A \exp (-iB \tan^{-1}(\sinh x))P^{(\alpha,\beta)}_n(i\sinh x),
\ee
with $\alpha=B-A-\frac{1}{2}$ and $\beta=-B-A-\frac{1}{2}$.\\
The superpotential
\be
W(x)=A\tanh x+iB\sech x
\ee
and the partner potential is
\be
V^{(A,B)}_{2,Scar II}(x)=-(B^2+A(A-1))\sech^2 x+iB(2A-1)\sech x\tanh x.
\ee
Thus the potential is translationally shape invariant under the translation of parameter $A\rightarrow A-1$.

In this case, the symmetry under the translation of parameters $B\longleftrightarrow (A+\frac{1}{2})$
has been already discussed in Ref. \cite{sl2c}. Here we mention only the results to get the consistency with the new rationally
extended case discussed in next section.

The bound state energy eigenvalues and the wavefunctions under the above parametric transformation are
given by
\be\label{dbs2}
E_n=-(B-n-\frac{1}{2})^2;\qquad n=0,1,2,...,n_{\mbox{max}}<B-\frac{1}{2},
\ee
and
\be
\psi^{(B\leftrightarrow A+\frac{1}{2})}_n(x)=N^{B\leftrightarrow A+\frac{1}{2}}_n (\sech x)^{B-\frac{1}{2}} \exp (-i(A+\frac{1}{2}) \tan^{-1}(\sinh x))P^{(\alpha,\beta)}_n(i\sinh x)
\ee
respectively, where the parameters $\alpha=A-B+\frac{1}{2}$ and $\beta=-A-B-\frac{1}{2}$.\\
The transimission $t(k)$ and reflection $r(k)$ amplitudes for the potential (\ref{scarfii_pt}) is in fact invariant under 
the transformation $B\longleftrightarrow A+\frac{1}{2}$, as can be seen from the expressions \cite{za} 
\be\label{scarftk}
t(k)=\frac{\Gamma(-A-ik)\Gamma(A+1-ik)\Gamma(-B-ik+\frac{1}{2})\Gamma(B-ik+\frac{1}{2})}{\Gamma(-ik)\Gamma(1-ik)\Gamma^2 (\frac{1}{2}-ik)}
\ee
and
\be
r(k)=t(k)\big[ i\cos (\pi A)\sin(\pi B)\sech (\pi k)+i\sin (\pi A)\cos (\pi B)\cosech (\pi k)\big].
\ee
The poles of the gamma functions $\Gamma(-A-ik)$ and $\Gamma(-B-ik+\frac{1}{2})$ give the exact bound state energy spectrums (\ref{scarfptee1}) 
and (\ref{dbs2}) respectively.

\subsubsection{Rationally extended $PT$ symmetric complex Scarf II potential}

The rationally extended $PT$ symmetric complex Scarf II potential \cite{bqr} isospectral to the conventional one (\ref{scarfii_pt}) 
is given by 
\ba\label{rescarf2}
V^{(A,B)}_{1,Scarf II}(x)&=&-(B^2+A(A+1))\sech^2 x+iB(2A+1)\sech x\tanh x\nonumber\\
&+&\frac{-2(2A+1)}{(-2iB\sinh x+2A+1)}+\frac{2[(2A+1)^2-B^2]}{(-2iB\sinh x+2A+1)^2}.
\ea 
The eigenfunctions in terms of $X_1$ exceptional Jacobi polynomials associated with this system are
\be
\psi^{(A,B)}_n(x)=N_{n,extd}\frac{(\sech x)^{A}\exp{\{-iB\tan^{-1}(\sinh x)\}}}{-2iB\sinh x+2A+1}\hat{P}^{(\alpha,\beta)}_{n+1}(i\sinh x).
\ee
Similar to the above two cases of real potentials, now we show that the parametric transformation $B\longleftrightarrow A+\frac{1}{2}$ 
 in the conventional $PT$ symmetric complex Scarf II potential leads to a new rationally extended
$PT$ symmetric Scarf II potential 
\ba\label{nres2}
V^{(A,B)}_{1,extd}(x)&=&-(B^2+A(A+1))\sech^2 x+iB(2A+1)\sech x\tanh x\nonumber\\
&+&\frac{-4B}{(-2i(A+\frac{1}{2})\sinh x+2B)}+\frac{8(B^2-(A+\frac{1}{2})^2)}{(-2i(A+\frac{1}{2})\sinh x+2B)^2},
\ea
whose bound state energy eigenvalues are same as given in Eq. (\ref{dbs2}).
The wave functions in terms of $X_1$ exceptional orthogonal Jacobi polynomials become
\be
\psi^{(B\leftrightarrow A+\frac{1}{2})}_n(x)=N^{(B\leftrightarrow A+\frac{1}{2})}_{n,extd}\frac{(\sech x)^{B-\frac{1}{2}}\exp{\{-i(A+\frac{1}{2})\tan^{-1}(\sinh x)\}}}{-2i(A+\frac{1}{2})\sinh x+2B}\hat{P}^{(\alpha,\beta)}_{n+1}(i\sinh x).
\ee
The superpotential corresponding to this new potential is
\ba
W(x)&=&(B-\frac{1}{2})\tanh x + i(A+\frac{1}{2})\sech x+2i (A+\frac{1}{2})\cosh x\nonumber\\
&\times & \bigg(\frac{1}{-2i(A+\frac{1}{2})\sinh x+2B-2}-\frac{1}{-2i(A+\frac{1}{2})\sinh x+2B}\bigg)
\ea
and the partner potential is
\ba
V^{(A,B)}_{2,extd}(x)&=&-((B-1)^2+A(A+1))\sech^2 x+i(B-1)(2A+1)\sech x\tanh x\nonumber\\
&+&\frac{-4(B-1)}{(-2i(A+\frac{1}{2})\sinh x+2B-2))}+\frac{8((B-1)^2-(A+\frac{1}{2})^2)}{(-2i(A+\frac{1}{2})\sinh x+2B-2))^2}.\nonumber\\
\ea

Similar GPT and Scarf I, this new rationally extended potential also generalizes to the potentials whose solutions are in terms of $X_m$
exceptional Jacobi polynomials given by
\ba\label{scarf23}
V^{(A,B)}_{m}(x)&=&(-B^2-A(A+1))\sech^2 x+iB(2A+1)\sech x \tanh x\nonumber\\
&+&2m(2A-m+2)+(2A-m+2)\nonumber\\
&\times &[(-2B)+(2A+2)i\sinh x]\frac{P^{(-\alpha,\beta)}_{m-1}(i\sinh x)}{P^{(-\alpha-1,\beta-1)}_{m}(i\sinh x)}\nonumber\\
&-&\frac{(2A-m+2)^2\cosh^2x}{2}\bigg (\frac{P^{(-\alpha,\beta )}_{m-1}(i\sinh x)}
{P^{(-\alpha-1,\beta-1 )}_{m}(i\sinh x)}\bigg )^2 \nonumber\\
\ea
and
\be\label{wfscarf23}
\psi^{(B\leftrightarrow A+\frac{1}{2})}_{n,m}(x) = N^{(B\leftrightarrow A+\frac{1}{2})}_{n,m,extd}\times \frac{(\sech x)^{A}\exp (-iB\tan^{-1}(\sinh x))}{P_{m}^{(-\alpha-1,\beta-1)}(i\sinh x)}\hat{P}_{n+m}^{(\alpha,\beta)}(i\sinh x). 
\ee 
This potential is also isospectral to the conventional one whose bound state energy eigenvalues are given by Eq. (\ref{dbs2}).
The above new potential is still SI under the translation of parameter $B\longrightarrow B-1$.

\section{The $so(2,1)$ algebra and its realizations}

In Ref. \cite{gtextd}, we have extended the works of Alhassid et al \cite{potalg1,potalg2,potalg3,potalg4,potalg5,potalg6,
potalg7} and obtained the modified generators $J_{\pm}$ and $J_3$
corresponding to $so(2,1)$ algebra for the rationally extended potentials
whose solutions are in the form of EOPs. After observing different parametric symmetries in the
conventional potentials it is interesting to see that the generators corresponding to these conventional
potentials are different with different Casimir operators. We have also shown that
these parametric symmetries generate a new set of rationally extended potentials whose solutions are
in the forms of exceptional Jacobi polynomials. In this section the generators of
the above algebra corresponding to these new rationally extended potentials are also obtained.

In this section, first we briefly review the $so(2,1)$ potential algebra
 and its unitary representations. This algebra consists of three generators $J_{\pm}$ and $J_{3}$ and satisfy the commutation relations
\be\label{commut}
[J_{+},J_{-}]=-2J_{3};\qquad [J_{3},J_{\pm}]=\pm J_{\pm}\,.
\ee
The differential realization of these generators corresponding to the conventional potentials \cite{potalg5} is given by
\ba\label{udre}
J_{\pm}&=&e^{\pm i\phi}\bigg[\pm\frac{\partial}{\partial x}
-\bigg( (-i\frac{\partial}{\partial \phi}\pm\frac{1}{2})F(x)
-G(x)\bigg)\bigg]\,,\nonumber \\
J_{3}&=&-i\frac{\partial}{\partial\phi}\,.
\ea
However, we find that these generators are not sufficient to explain the
spectrum of the rationally extended SI potentials. Hence,
we have constructed the $so(2,1)$ algebra by modifying $J_{\pm}$ with the inclusion of a new operator, $U(x,-i\frac{\partial}{\partial \phi}\pm\frac{1}{2})$
 \cite{gtextd} as,
\be\label{extdre}
J_{\pm}=e^{\pm i\phi}\bigg[\pm\frac{\partial}{\partial x}
-\bigg( (-i\frac{\partial}{\partial \phi}\pm\frac{1}{2})F(x)
-G(x)\bigg)-U(x,-i\frac{\partial}{\partial \phi}\pm\frac{1}{2})\bigg]\,,
\ee
and keeping the generator $J_{3}$ unchanged.\\
Here $F(x)$, $G(x)$ and $U(x,-i\frac{\partial}{\partial\phi}\pm\frac{1}{2})$ are two functions and a functional operator respectively.

In order to satisfy the $so(2,1)$ algebra (\ref{commut}) by these new generators $J_{\pm}$ and $J_{3}$, the following
restrictions on the functions $F(x)$, $G(x)$ and $U(x, k\pm\frac{1}{2})$
\ba\label{rest1}
\frac{d}{dx}F(x)+F^2(x)=1;\qquad \frac{d}{dx}G(x)+F(x)G(x)=0;
\ea
and
\ba\label{rest2}
\bigg[U^2(x,k-\frac{1}{2})-\frac{d}{dx}U(x,k-\frac{1}{2})+2U(x,k-\frac{1}{2})\bigg( F(x)(k-\frac{1}{2})-G(x)\bigg)\bigg]-\nonumber\\
\bigg[U^2(x,k+\frac{1}{2})+\frac{d}{dx}U(x,k+\frac{1}{2})
+2U(x,k+\frac{1}{2})\bigg( F(x)(k+\frac{1}{2})-G(x)\bigg)\bigg]=0\,
\ea
are required.

Here we note that the Eq. (\ref{rest1}) is the same as for the conventional potentials
\cite{potalg5} while an additional condition (\ref{rest2}) appears
due to the presence of the extra term
$U(x,-i\frac{\partial}{\partial \phi}\pm\frac{1}{2})$ in $J_{\pm}$.

The Casimir operator, for the $so(2,1)$ algebra, in terms of the above
generators is given by
\be\label{cas}
J^2=J^2_{3}-\frac{1}{2}(J_{+}J_{-}+J_{-}J_{+})
= J^2_{3}\mp J_{3}-J_{\pm}J_{\mp}\,.
\ee
For the bound states, the basis for an irreducible representation of
extended $so(2,1)$ is characterized by
\be
J^2\ket{j,k}=j(j+1)\ket{j,k};\qquad J_{3}\ket{j,k}=k\ket{j,k}\,,
\ee
and
\be
J_{\pm}\ket{j,k}=[-(j\mp k)(j\pm k+1) ]^{\frac{1}{2}}\ket{j,k\pm1}\,.
\ee
Using (\ref{extdre}), the differential realization of the Casimir operator
in terms of $F(x)$, $G(x)$ and $U(x, J_{3}-\frac{1}{2})$ is given by
\ba\label{dcasmir}
J^2&=&\frac{d^2}{dx^2}+\bigg(1-F^2(x)\bigg)(J^2_{3}-\frac{1}{4}) - 2\frac{dG(x)}{dx}(J_{3})-G^2-\frac{1}{4}\nonumber\\
&-&\bigg[U^2(x,J_{3}-\frac{1}{2})+\bigg(\big(J_{3}-\frac{1}{2}\big) F(x)-G(x)\bigg) U(x,J_{3}-\frac{1}{2})\nonumber\\
&+&U(x,J_{3}-\frac{1}{2})\bigg(\big( J_{3}
-\frac{1}{2}\big) F(x)-G(x)\bigg)-\frac{d}{dx}U(x,J_{3}-\frac{1}{2})\bigg]\,,
\ea
 and the basis \ket{j,k} in the form of function is given as
 \be\label{basis}
 \ket{j,k}=\psi_{jk}(x,\phi)\simeq \psi_{jk}(x)e^{ik\phi}\,.
 \ee
The functions (\ref{basis}) satisfy the Schro\"odinger equation
\be\label{sch}
\bigg[-\frac{d^2}{dx^2}+V_{k}(x)\bigg]\psi_{jk}(x)=E\psi_{jk}(x)\,,
\ee
where $V_{k}(x)$ is one parameter family of $k$-dependent potentials given by
\ba\label{kpot}
V_{k}(x)&=&(F^2(x)-1)(k^2-\frac{1}{4})+2k\frac{d}{dx}G(x)+G^2(x)\nonumber\\
&+&\bigg[U^2(x,k-\frac{1}{2})+2\bigg(\big( k-\frac{1}{2}\big) F(x)-G(x)\bigg) U(x,k-\frac{1}{2})\nonumber\\
&-&\frac{d}{dx}U(x,k-\frac{1}{2})\bigg]\,,
\ea
and the corresponding energy eigenvalues are given by
\be\label{eng}
E_{j}=-\big(j+\frac{1}{2}\big)^2\,.
\ee
By replacing $U(x,k\pm\frac{1}{2})\Rightarrow  U(x,m, k\pm\frac{1}{2})$, the above realizations are also suitable 
for the one parameter family of rationally extended SI potentials associated with $X_m$ exceptional orthogonal polynomials.
For a check, if we put $m=0$ i.e. $U(x,0,k-\frac{1}{2})=0$ in the above Eq. (\ref{kpot}), we get the expressions for
corresponding conventional potentials discussed  in Ref. \cite{potalg5}.

Thus the Hamiltonian in terms of the Casimir operator of $so(2,1)$ algebra is
given by
\be\label{ham}
H=-\big(J^2+\frac{1}{4}\big)\,.
\ee
 It may be noted that the $so(2,1)$ algebra (\ref{extdre}) with the modified
generators satisfies the same unitary representation as satisfied by the
generators corresponding to the usual potentials \cite{potalg5}. Here we
discuss the unitary representation of $so(2,1)$ algebra corresponding to
the discrete principal series $D^{+}_{j}$ for which $j<0$ i.e.,
\be\label{uniprep1}
 k=-j+n;\quad n=0,1,2,...,.
\ee
Thus the energy eigenvalues (\ref{eng}) corresponding to this series will be
\be\label{eng1}
E_{n}=-\big(n-(k-\frac{1}{2})\big)^2\,.
\ee

\subsection{ Generalized P\"oschl Teller potential}

For GPT potential, the spectrum under the change of parameters $B\leftrightarrow A+\frac{1}{2}$ can 
be obtained by choosing 
\be
 F(x)=\coth x;\qquad G(x)=(A+\frac{1}{2})\cosech x\,.
\ee
For new extended GPT potential whose bound states are in terms of $X_1$ EOPs as given by (\ref{gptex1}), we also need 
to choose $U(x,k\pm\frac{1}{2})$ as
\be
U(x,k\pm\frac{1}{2})=\bigg(\frac{(2A+1)\sinh x}{(2A+1) \cosh x-2(k\pm\frac{1}{2})-1}
-\frac{(2A+1)\sinh x}{(2A+1)\cosh x-2(k\pm\frac{1}{2})+1}\bigg); 
\ee
with $A > k > 0$. On substituting these functions in (\ref{kpot}), we get the expression for new rationally
extended GPT potential given in equation (\ref{gptex1}) with the parameter $B$ is replaced by $k$.
The energy eigenvalues of this extended potential are same as that of
conventional one (i.e they are isospectral) (\ref{eng1}) and the
associated wavefunctions $\psi_{jk}(x)$ (\ref{sch}) are given in terms of $X_1$ exceptional Jacobi polynomials.

The new rationally extended potentials (\ref{gptpot3}) corresponding to the $X_m$  case can be obtained by assuming  
\ba
U(x,k\pm\frac{1}{2})&\Rightarrow &U(x,m,k\pm\frac{1}{2}) \nonumber \\
&=& \frac{(m-2A-2)\sinh x}{2}\times\bigg[\frac{P^{((k\pm\frac{1}{2})-A,-(k\pm\frac{1}{2})-A-1)}_{m-1}(\cosh x)}{P^{(k\pm\frac{1}{2}-A-1,-(k\pm\frac{1}{2})A-2)}_{m}(\cosh x)}\nonumber\\
&-&\frac{P^{((k\pm\frac{1}{2})-A-1,-(k\pm\frac{1}{2})-A)}_{m-1}(\cosh x)}{P^{(k\pm\frac{1}{2}-A-2, -(k\pm\frac{1}{2})-A-1)}_{m}(\cosh x)}\bigg]\,,
\ea
where $P^{(\alpha, \beta)}_{m}(\cosh x)$ is conventional
Jacobi polynomials. The energy eigenvalues will be same as given in (\ref{eng1}).\\

For $m=0$, the function i.e $U(x,m,k\pm\frac{1}{2})=U(x,0,k-\frac{1}{2})=0$, we get the
conventional GPT potential. 


\subsection{$PT$ symmetric complex Scarf-II potential}

In this section, we consider the new conventional and rationally extended $PT$ symmetric complex
Scarf II potential obtained in section $2$ after the transformation of parameters $B\longleftrightarrow A+\frac{1}{2}$.
For these complex potential we use extended $sl(2,\mathbb{C})$ potential algebra.
 In this algebra at least one of functions  $F(x)$, $G(x)$ and
$U(x,k\pm\frac{1}{2})$ must be complex
and satisfy Eqs. (\ref{rest1}) and (\ref{rest2}).

For the conventional $PT$ symmetric complex Scarf II potential which is invariant under $B\leftrightarrow A+\frac{1}{2}$, we consider the
functions
\be
F(x)=\tanh x;\qquad G(x)=i(A+\frac{1}{2}) \sech x. 
\ee
In addition to these functions new rationally extended complex Scarf II potential (\ref{nres2}) associated with $X_1$ EOPs are obtained by 
defining
\ba
U(x,k-\frac{1}{2})&\Rightarrow& U(x,1,k-\frac{1}{2})\nonumber\\
&=& \bigg[ \frac{2i(A+\frac{1}{2})\cosh x}{(-2i(A+\frac{1}{2})\sinh x+2k-2)}-\frac{2i(A+\frac{1}{2})\cosh x}{(-2i(A+\frac{1}{2})\sinh x+2k)}\bigg]\,.
\ea
On putting all these functions $F(x)$, $G(x)$ and $U(k-\frac{1}{2})$ in (\ref{kpot}), we get the expression for 
new rationally extended $PT$ symmetric potential (which is on the full-line $-\infty\leq x\leq\infty$) (\ref{nres2}) with the 
parameter $B$ is being replaced by $k$.\\
The energy eigenvalues for this extended complex potential are real and are
the same as that of the conventional one given by
\be\label{enscarf}
E_n =-(n-k+\frac{1}{2})^2;\quad n= 0,1, . . . ,n_{max};
\qquad n_{max} < (k-\frac{1}{2})\,,
\ee 
and the associated wavefunctions $\psi_{jk}(x)$ (\ref{sch}) are given in terms of $X_1$ exceptional Jacobi polynomials.
The new rationally extended complex Scarf II potentials (\ref{scarf23}) having same energy eigenvalues (\ref{enscarf}) and associated with the $X_m$ EOPs 
are obtained by assuming 
\ba
U(x,k\pm\frac{1}{2})&\Rightarrow &U(x,m,k\pm\frac{1}{2}) \nonumber \\
&=& \frac{(m-2B-2)i\cosh x}{2}\times\bigg[\frac{P^{(k\pm\frac{1}{2}-A,-(k\pm\frac{1}{2})-A-1)}_{m-1}(i\sinh x)}{P^{(k\pm\frac{1}{2}-A-1,-(k\pm\frac{1}{2})-A-2)}_{m}(i\sinh x)}\nonumber\\
&-&\frac{P^{(k\pm\frac{1}{2})-A-1,-(k\pm\frac{1}{2})-A)}_{m-1}(i\sinh x)}{P^{(k\pm\frac{1}{2}-A-2, -(k\pm\frac{1}{2})-A-1)}_{m}(i\sinh x)}\bigg]\,.
\ea
For $m=0$, the function $U(x,m,k\pm\frac{1}{2})$ becomes zero, hence we obtain the usual case of $sl(2,\mathbb{C})$ and the corresponding conventional 
complex Scarf II potential. On the other hand for $m=1$, we recover our results corresponding to the $X_1$ case as discussed above.

\subsection{The $iso(2,1)$ algebra and trigonometric  Scarf  potential}

In Ref. \cite{gtextd}, we have discussed the potential algebra approach to the rationally extended GPT and $PT$ 
symmetric complex Scarf II potentials only. Following the works of Levai \cite{levai}, we are now able to obtain the 
bound state spectrums of rationally extended SI Scarf I potential. In this section first we obtain the spectrums of rationally 
extended Scarf I potentials (\ref{extsc_con_pot}) and then we consider the new rationally extended Scarf I potentials 
obtained after the parametric transformation $B\longleftrightarrow A-\frac{1}{2}$. 

For these potentials, the above $so(2,1)$ algebra is not suitable. 
The algebra corresponding to this potential is obtained by multiplying the generators $J_{\pm}$ 
of $so(2,1)$ algebra with an imaginary number $i$, thus the resulting potential algebra for 
this potential is $iso(2,1)$. The modified generators $J_{\pm}$ corresponding to this
algebra are given as
\be\label{extdreiso}
J_{\pm}=ie^{\pm i\phi}\bigg[\pm\frac{\partial}{\partial x}
+\bigg( (-i\frac{\partial}{\partial \phi}\pm\frac{1}{2})F(x)
-G(x)\bigg)+U (x,-i\frac{\partial}{\partial \phi}\pm\frac{1}{2})\bigg ]\,.
\ee
Similar to the $so(2,1)$ case, to satisfy $iso(2,1)$ algebra by these generators $J_{\pm}$ and $J_{3}$, the commutation relations (\ref{commut}) have to be satisfied. This requirements provide following
restrictions on the functions $F(x)$, $G(x)$ and $U(x, k\pm\frac{1}{2})$
\ba\label{restiso1}
\frac{d}{dx}F(x)-F^2(x)=1;\qquad \frac{d}{dx}G(x)-F(x)G(x)=0;
\ea
and
\ba\label{restiso2}
\bigg[U^2(x,k+\frac{1}{2})-\frac{d}{dx}U(x,k+\frac{1}{2})+2U(x,k+\frac{1}{2})\bigg( F(x)(k+\frac{1}{2})-G(x)\bigg)\bigg]-\nonumber\\
\bigg[U^2(x,k-\frac{1}{2})+\frac{d}{dx}U(x,k-\frac{1}{2})
+2U(x,k-\frac{1}{2})\bigg( F(x)(k+\frac{1}{2})-G(x)\bigg)\bigg]=0\,.
\ea
Here also we note that Eq. (\ref{restiso1}) is the same as for the usual potentials
\cite{levai} while an extra Eq. (\ref{restiso2}) appears
due to the presence of the extra term
$U(x,-i\frac{\partial}{\partial \phi}\pm\frac{1}{2})$.\\
 Using (\ref{extdreiso}), the differential realization of the Casimir operator
in terms of $F(x)$, $G(x)$ and $U(x, J_{3}+\frac{1}{2})$ is given by
\ba\label{ciso}
J^2&=&-\frac{d^2}{dx^2}+(1+F^2(x))\big (J^2_{3}-\frac{1}{4}\big) - 2\frac{dG(x)}{dx}(J_{3})+G^2-\frac{1}{4}\nonumber\\
&+&U^2(x,J_{3}+\frac{1}{2})+\bigg(\big(J_{3}+\frac{1}{2}\big) F(x)-G(x)\bigg) U(x,J_{3}+\frac{1}{2})\nonumber\\
&+&U(x,J_{3}+\frac{1}{2})\bigg(\big( J_{3}
+\frac{1}{2}\big) F(x)-G(x)\bigg)-\frac{d}{dx}U(x,J_{3}+\frac{1}{2})\,.
\ea
 Thus the Hamiltonian in terms of the Casimir operator of $iSO(2,1)$ algebra is
given by
\be\label{hiso}
H=\big(J^2+\frac{1}{4}\big)\,.
\ee
The unitary representation of $iso(2,1)$ algebra will be same as of $so(2,1)$.

The spectrums of the conventional Scarf I potential (\ref{scarf1}) are obtained by considering \cite{levai}
\be
 F(x)=\tan x;\qquad G(x)=B\sec x;\quad -\frac{\pi}{2}<x<\frac{\pi}{2};\quad 0 < B<k-\frac{1}{2}\,, \nonumber
\ee
and for the rationally extended Scarf I potential (\ref{extsc_con_pot}) we also choose
\ba
U(x,k+\frac{1}{2})&=&\bigg[\frac{-2B\cos x}{(-2B \sin x+2(k+\frac{1}{2})-1)}
+\frac{2B\cos x}{(-2B \sin x+2(k+\frac{1}{2})+1)}\bigg]\,,\nonumber\\
\ea
so that the above conditions (\ref{restiso1}) and (\ref{restiso2}) are satisfied.
On substituting these functions in (\ref{ciso}), we get the rationally
extended Scarf I potential (\ref{extsc_con_pot}) with the parameter $A$ is replaced $k+\frac{1}{2}$. 
The energy eigenvalues of this extended potential are isospectral to their conventional counterpart
and are given by Eq. (\ref{hiso}) and (\ref{eng1}) i.e.
\be
E_n = (j+\frac{1}{2})^2.
\ee
The associated wavefunctions of this potentials are given in terms of $X_1$ exceptional Jacobi polynomial.\\
Similar to the rationally extended GPT and $PT$ symmetric Scarf II potentials \cite{gtextd} the extended Scarf I potential
whose solutions are in terms of $X_m$ EOPs can be obtained by defining
\ba
U(x,k\pm\frac{1}{2})&\Rightarrow &U(x,m,k\pm\frac{1}{2}) \nonumber \\
&=& \frac{(2B+m-1)\cos x}{2}\times\bigg[\frac{P^{(-\alpha,\beta )}_{m-1}(\sin x)}{P^{(-\alpha-1,\beta)}_{m}(\sin x)}\nonumber\\
&-&\frac{P^{(-\alpha-1,\beta+1)}_{m-1}(\sin x)}{P^{(-\alpha-2, \beta)}_{m}(\sin x)}\bigg]\,,
\ea
with $\alpha=k\pm\frac{1}{2}-B-\frac{1}{2}$ and $\beta=k\pm\frac{1}{2}+B-\frac{1}{2}$. 
For $m=0$, $U(x,k\pm\frac{1}{2})\Rightarrow U(x,0,k\pm\frac{1}{2})=0$, Eq. (\ref{ciso}) produces the conventional Scarf I potential.

 After transforming the parameters $B\leftrightarrow A-\frac{1}{2}$, the conventional Scarf I potential (\ref{scarf1})
 remain invariant which can be obtained by using 
 \be
 F(x)=\tan x; \qquad G(x)=(A-\frac{1}{2})\sec x.
 \ee
 For the new rationally extended Scarf I potential (\ref{nscarf}), in addition to $F(x)$ and $G(x)$ we define
 \ba
U(x,k+\frac{1}{2})&=&\bigg[\frac{(2A-1)\cos x}{2(k+\frac{1}{2})+1-(2A-1)\sin x}
-\frac{(2A-1)\cos x}{2(k+\frac{1}{2})-1-(2A-1)\sin x}\bigg]\,.\nonumber\\
\ea
The potentials corresponding to the $X_m$ case (\ref{nscarfxm}) are obtained by constructing
 \ba
U(x,k\pm\frac{1}{2})&\Rightarrow &U(x,m,k\pm\frac{1}{2}) \nonumber \\
&=& \frac{(2A+m-2)\cos x}{2}\times\bigg[\frac{P^{(-(k\pm\frac{1}{2})-A,k\pm\frac{1}{2}+A-1)}_{m-1}(\sin x)}{P^{(-(k\pm\frac{1}{2})+A-1,k\pm\frac{1}{2}+A-1)}_{m}(\sin x)}\nonumber\\
&-&\frac{P^{(-(k\pm\frac{1}{2})+A-1,k\pm\frac{1}{2}+A)}_{m-1}(\sin x)}{P^{(-(k\pm\frac{1}{2})+A-2, k\pm\frac{1}{2}+A-1)}_{m}(\sin x)}\bigg]\,.
\ea
 
Using these functions in Eq. (\ref{ciso}), we get the expression for new rationally extended Scarf I potentials Eq. (\ref{nscarfxm})
with the parameters $B$ is replaced by $k$.

\section{Summary and discussion}

In this work we have discussed the different parametric symmetries in conventional
real as well as $PT$ symmetric complex potential. It has been shown that the above
symmetry provides a new set of bound states with new superpotentials.
 As examples we consider three conventional potentials namely GPT, Scarf I and $PT$ symmetries
Scarf II potentials and obtained corresponding new potentials which are invariant
under the above symmetries with completely different bound states. The rationally extended potentials
corresponding to these new potentials are obtained whose solutions are in terms of EOPs. New ES potentials which are
isospectral to the conventional
potentials are obtained for
all three cases.

Further we have studied the effect of these parametric symmetries powerful technique of group theoretic method
in which the Hamiltonian for the new conventional as well as rationally extended GPT,
Scarf I and $PT$ symmetric  Scarf-II systems are expressed purely in terms of the
modified Casimir operator of $so(2,1)$, $iso(2,1)$ and $sl(2,\mathbb{C})$
groups respectively.

{\bf Acknowledgments}

   BPM acknowledges the financial support from the Department of Science and Technology (DST), Govt. of India under SERC project sanction
grant No. SR/S2/HEP-0009/2012. AK wishes to thank Indian National Science Academy (INSA) for the award of INSA senior 
scientist position at Savitribai Phule Pune University.
NK acknowledges the financial support
from BHU Varanasi in the form of CRET fellowship.

\end{document}